\documentclass[twocolumn,showpacs,preprintnumbers,amsmath,amssymb]{revtex4}

\usepackage{epsfig,psfrag}
\usepackage{dcolumn}
\usepackage{bm}
\usepackage{graphicx}
\usepackage{color}

\begin{document}
\title{Path-integral Monte Carlo simulations for interacting
 few-electron quantum dots with spin-orbit coupling  }
\author{Stephan Weiss and R. Egger}
\affiliation{Institut f\"ur Theoretische Physik,
Heinrich Heine Universit\"at, D-40225 D\"usseldorf, Germany
}
\date{\today}
\begin{abstract}
We develop path-integral Monte Carlo simulations for
a parabolic two-dimensional (2D) quantum dot 
containing $N$ interacting electrons in the presence of 
Dresselhaus and/or Rashba spin-orbit couplings.  
Our method solves in a natural way the spin contamination problem
and allows for numerically exact finite-temperature results at
weak spin-orbit coupling.  For $N<10$ electrons, we present data for  
the addition energy, the particle density, 
and the total spin $S$ in the Wigner molecule regime
of strong Coulomb interactions. We identify magic numbers at $N=3$ and 
$N=7$ via a peak in the addition energy.  These magic numbers
differ both from weak-interaction and classical predictions,
and are stable with respect to (weak) spin-orbit couplings.
\end{abstract}
\maketitle

\section{Introduction}
\label{intro}

Interacting few-body quantum systems have attracted a lot of attention
over the past decades. 
In that respect, quantum dots (QDs) 
continue to be of fundamental interest to several fields, 
e.g., spintronics, nanoelectronics, and quantum computing. QDs are
small solid-state devices, typically 
containing a few up to several hundred electrons
confined in all space directions \cite{reimann,alhassid}.
They can be fabricated and 
studied using different approaches and
materials \cite{kouwenhoven,ashori,tarucha1,tarucha2,lindemann},
and control over both the charge and the spin degree
of freedom of the confined electrons has been reported 
in experiments.  Here, we address QDs as realized in 
two-dimensional (2D) semiconductor devices, e.g., 
in ultraclean 2D electron gases.  
We consider the case of parabolically confined individual QDs,
which is quite appropriate
in most practical cases \cite{reimann}.
QDs offer the possibility to tune Coulomb correlations among
electrons via external gates.   

Spin properties of quantum dots have recently entered the focus of research 
\cite{Nitta1,schapers,grundler,Nitta2,koenemann}, 
in particular as they are central both to quantum computation \cite{loss}
and to spintronics \cite{dassarma}.
Here spin-orbit (SO) terms have to be taken into
account, coupling the spin dynamics to the orbital motion.
In general, SO coupling is a relativistic effect, and
appears to second order in the fine structure constant.
In most materials of interest, two main mechanisms may be
distinguished, namely Rashba \cite{Rashba}
and Dresselhaus \cite{Dresselhaus} SO couplings.
The Rashba SO coupling strength $\alpha_R$ is due to the surface inversion
asymmetry present in the confinement to a 2D electron gas, and
therefore can be tuned  
by external gates \cite{Nitta1}. The Dresselhaus coupling $\alpha_D$
is generally not tunable
but can be important if the host crystal has no bulk inversion symmetry,
e.g., in zincblende semiconductors. For simplicity, we 
only consider the linear Dresselhaus term and neglect various 
additional spin-orbit contributions, see
Ref.~\cite{destefani2} for an extended discussion.  These 
contributions can in principle be taken into account within our
approach. Furthermore,
we focus on the case of vanishing magnetic field.

In this work, we investigate the behavior of a closed parabolic
few-electron QD in the presence of spin-orbit couplings, containing up 
to $N=9$ interacting electrons.
While the numerically exact method employed here allows to study arbitrary 
interactions in principle, it is probably most useful 
in the regime of intermediate-to-strong Coulomb interactions, 
where a `Wigner molecule' 
\cite{Egger,Egger2,Filinov,Reusch,hausler,harting,rontani}
is formed.
In terms of the standard electron gas parameter $r_s$, the
studied interactions are around $r_s \approx 5$ to 10.
Then a clear tendency towards Wigner crystallization can be observed, but
different spatial ```shells'' are not yet locked relative to each 
other as happens for very large $r_s$ \cite{Filinov}.
In fact, while for $r_s\to\infty$, a completely classical 
situation is encountered \cite{lozovik,bedanov}, quantum
effects still play a major role for the `incipient' Wigner molecule
of interest here.  In such a case, many standard calculational tools,
e.g., exact diagonalization \cite{mikhailov},
the Hartree-Fock approximation, the 
 fixed-node \cite{pederiva} or variational
\cite{harju} Monte Carlo approach, or
density functional theory 
can meet  various difficulties (like artificial symmetry breakings)
or require explicit justification, see
Ref.~\cite{reimann} for a review.  
In that situation,
finite-temperature path-integral Monte Carlo (PIMC)
simulations represent an attractive alternative scheme.
The case of no spin-orbit coupling has been studied using PIMC in
Refs.~\cite{Egger,Filinov,Reusch,harting}.

Before describing our PIMC scheme and the ensuing results, let us
first discuss previous theoretical approaches to the physics
of QDs in the presence of SO couplings.
To study their effect on single-particle 
energies, one may set up perturbation theory for small SO couplings.
Due to the linear dependence on momentum in Eq.~(\ref{modHamSO}) below,
perturbation theory starts at second order and  
gives a quadratic decrease of single-particle energies with 
increasing Rashba coupling $\alpha_R$ (or Dresselhaus coupling $\alpha_D$)
\cite{gogolin}. 
Single-particle energy level crossings induced 
by the SO coupling have been discussed in Refs.~\cite{destefani2,vos,kuan} 
as a function of an applied magnetic field.
Here, we are mainly concerned with many-body effects due to the
Coulomb interaction.
For two electrons,  exact diagonalization studies
have been carried out for rather strong Rashba couplings
and weak interactions, as appropriate for InSb dots 
\cite{chakraborty,destefani1}.  
Energy spectra were examined, 
a jump in the magnetization of the dot as a function
of magnetic field  was found \cite{chakraborty},
and a favoring of exchange over direct interactions 
as a consequence of SO interactions was discussed \cite{destefani1}.
Governale \cite{governale} has employed spin-density functional
theory for $N\leq 16$. He found that a very strong SO coupling leads to new 
peaks and/or the suppression of Hund's rule peaks otherwise present in the 
addition energy spectrum. An additional in-plane magnetic field
was argued to imply paramagnetic behavior. 
Finally, in Ref.~\cite{emperador}, several approximate schemes have been
employed to study SO effects in weakly interacting
QDs with $N\approx 11$ to 13 electrons.
Here we provide results
for $N<10$ electrons with strong Coulomb interactions and SO couplings, 
where exact diagonalization techniques may not apply anymore.  
One should also note that for $r_s\approx 5$ to 10, spin effects are
very important but necessitate an essentially exact treatment.
For a recent comparison of QMC data to density functional
theory in that respect, see Ref.~\cite{baranger}.

In the absence of spin-orbit couplings, PIMC simulations
for QDs suffer from 
two well-known problems, namely the fermionic sign problem
and the spin contamination problem \cite{Egger}.
(The first problem can be relieved to some extent by the
multilevel blocking algorithm \cite{Egger2}. Here we 
restrict ourselves to a simpler ``brute-force'' approach.)
The spin contamination problem arises for $\alpha_R=\alpha_D=0$ because 
both the total QD spin $S$ and its $z$ component $S_z$
are good quantum numbers. Now $S_z$ is in practice fixed during the 
simulation since there are no
spin flip terms in the hamiltonian. At finite temperature, 
one then arrives at the undesirable situation where states
with different $S$ 
but the same $S_z$ contribute to 
the simulation.  This considerably complicates data analysis  and
represents a well-known problem affecting also other schemes, e.g., 
Hartree-Fock calculations.
In the presence of spin-orbit coupling, however,  neither $S$ nor $S_z$ are 
good quantum numbers and the full space of all $\{ S_z, S\}$ becomes 
accessible. With increasing SO couplings, we find that the
sign problem worsens exponentially, restricting the 
applicability of our approach to weak SO couplings.
Since in applications, SO effects are usually weak, however, this
restriction is not too severe.

In order to eliminate the spin contamination problem in 
the limit of zero SO coupling,
we may study a few finite but small values for the SO couplings, and then
extrapolate $\alpha_{R/D}\to 0$.  This allows to reliably
compute, for instance, the addition spectrum of the dot, where we find
peaks (corresponding to stability islands of these Wigner molecules)
for $N=3$ and $N=7$ electrons.
Furthermore, we  compute the dependence of the 
spin state, $\langle {\bf S}^2 \rangle=S(S+1),$ as a function
of particle number $N$, where
${\bf S}$ is the total spin operator.

The structure of the paper is as follows. After introducing the
model in Section \ref{sec:Model} we derive the short-time propagator  for 
interacting fermions in a parabolic QD subject to either 
Rashba or Dresselhaus SO coupling (or both of them at the same time),
and discuss the numerical scheme in some detail. 
Numerical results are presented in Section 
\ref{sec:Results}, and we conclude in Section \ref{conc}.
Throughout the paper, we put $\hbar=1$.

\section{Model and method}
\label{sec:Model}

We study the $N$-electron hamiltonian describing a closed parabolic
dot in a 2D electron gas, 
\begin{equation}
H=\sum\limits_{i=1}^{N}\left(\frac{{\bf p}_i^2}{2m^*}+
\frac{m^*\omega_0^2}{2}{\bf r}_i^2\right)+\sum\limits_{i<j}\frac{e^2/\kappa}
{|{\bf r}_i-{\bf r}_j|}+\sum\limits_{i=1}^{N}H_{SO}^{(i)},
\label{modHam}
\end{equation}
where $m^*$ is the effective electron mass and $\omega_0$ is the oscillator 
frequency. With $i=1,\ldots,N$, the vectors
${\bf p}_i,{\bf r}_i$ denote the 2D momenta and space coordinates of 
all $N$ electrons.  The Coulomb potential among the electrons contains
screening effects of the host material via the dielectric constant,
$\kappa$.
Measuring energies (lengths) in units of $\omega_0$ 
 ($\l_0=1/\sqrt{m^*\omega_0}$), a dimensionless Coulomb interaction 
parameter is given by $\lambda=e^2/(\kappa l_0\omega_0)$.
For common host materials, 
 $l_0$ is in the range of few up to hundreds of 
$nm$. The confinement energy
$\hbar\omega_0$ is then typically between $0.1$  up to a few $meV$, which 
allows 
directly to determine the strength of the Coulomb interaction $\lambda$.  
Typical $\lambda$ currently realized experimentally are between 0.5 and 5,
the latter value \cite{ashori} being already quite close to the regime
studied in our paper.   
For $\lambda> 1$,  Coulomb repulsion starts to dominate over 
the kinetic energy, and electrons spatially arrange on shells. 
This ``Wigner molecule'' regime \cite{Egger} is studied in this work.

Let us then address the SO couplings considered here.
Typically, two different SO couplings are of paramount importance
in semiconductor heterostructures, namely
the (linear) Dresselhaus and the Rashba SO 
coupling. 
We allow for both types and consider the spin-orbit
hamiltonian (of the $i$th particle)
\begin{equation}\label{modHamSO}
H_{SO}=\alpha_R(p_x\sigma^y-p_y\sigma^x)
+\alpha_D(p_x\sigma^x-p_y\sigma^y)
 =  {\bf p}\cdot A\cdot\vec\sigma,
\end{equation}
where the standard Pauli matrices
$\vec\sigma=(\sigma^x,\sigma^y)$ 
act in spin space and 
\[
A= \left( \begin{array}{cc} 
\alpha_D &\alpha_R\\ -\alpha_R&-\alpha_D \end{array}
\right).
\]
Note that both types of SO coupling 
can be transformed into each other by a
unitary transformation.
Hence the spectra for $\alpha_D=0 ~ (\alpha_R\neq0)$
 and $\alpha_R=0 ~ (\alpha_D\neq 0)$ coincide (for zero magnetic
field), see also Sec. \ref{sec:Results} below.   

Remarkably, even on the single-particle level,
there is no closed solution to the Schr\"odinger equation 
in a parabolic potential subject to SO coupling of any kind. 
We mention in passing that for a cylindrical box, the single-particle problem
has been solved analytically in Ref.~\cite{gogolin}.
PIMC simulations provide a powerful tool to extract
numerically exact results for this few-body interacting quantum
system.  One starts by discretizing imaginary time ($0\leq t < \beta=
1/k_B T$) into sufficiently short time intervals $\tau=\beta/P$,
where $P$ is the (integer) Trotter number.
For $t=\tau$, an approximate but accurate
short-time propagator can be constructed via the Trotter-Suzuki
decomposition.  To that end, we split $H$ into the noninteracting
part $H_0$ (including the SO couplings) and the remaining interaction part
($\sim \lambda$).  Assuming that the spin dynamics is slow on the
timescale $\tau$, some algebra along the lines of Ref.~\cite{zinn}
then gives the short-time single-particle
propagator under $H_0$ in the form
\begin{eqnarray}
\label{single}
&& \langle {\bf r}_2,\sigma_2|e^{-\tau H_0}|
{\bf r}_1,\sigma_1 \rangle =
\frac{e^{\tau E_0}}{2\pi l_0^2\sinh(\omega_0\tau)}
\\ \nonumber &\times& 
e^{-S_0[ {\bf r}_2,{\bf r}_1]}  \
\left\langle \sigma_2\left|
e^{-im^* ({\bf r}_2-{\bf r}_1)
\cdot A\vec\sigma}\right|\sigma_1\right\rangle
\end{eqnarray}
where $E_0=m^*(\alpha_R^2+\alpha_D^2)$.  
The electron is described by its position ${\bf r}$ and the
$z$-component $\sigma/2$ of the spin.
Furthermore, $S_0$ denotes the standard 2D oscillator action 
\[
S_0[{\bf r}_2,{\bf r}_1]=\frac{
\left({\bf r}_1^2+{\bf r}_2^2\right)\cosh(\omega_0\tau)-2{\bf r}_1\cdot
{\bf r}_2}
{2l_0^2\sinh(\omega_0\tau)}.
\]
Note that in the absence of SO couplings, the exact propagator of the
 harmonic oscillator is reproduced.  
For small coupling constants $\alpha_{D,R}$ and sufficiently 
short $\tau,$
Eq.~(\ref{single}) represents a very accurate short-time approximation
to the single-particle propagator. 
The spin part in the propagator (\ref{single})  can be written as
\begin{eqnarray*}
&&
\left\langle \sigma_2|
e^{-im^* ({\bf r}_2-{\bf r}_1)
\cdot A\vec\sigma}|\sigma_1\right\rangle
= \cos(a) \delta_{\sigma_1, \sigma_2} \\
&& -i \  \frac{\sin (a)}{a} \
( a_x+i\sigma_1a_y) \delta_{\sigma_1,-\sigma_2},
\end{eqnarray*}
where ${\bf a}=(a_x,a_y)= m^* A^T ({\bf r}_2-{\bf r}_1)$ and $a=|{\bf a}|$.
Using Eq.~(\ref{single}), the many-body 
propagator follows in the form of $N\times N$ Slater determinants. 
Let us then denote the coordinate and the spin $\sigma$
 of the $i$th electron on time slice $n$ (where
$1\leq n \leq P$) as ${\bf r}_{in}$ and $\sigma_{in}$, respectively.
With respect to the time direction, we have periodic
boundary conditions.
Including Coulomb interactions, we then obtain the many-particle 
partition function for given discretization $\tau=\beta/P$ 
in the form 
\begin{eqnarray}
\label{partfun}
Z &=& 
 \sum_{\{\sigma_{jn}=\pm\}} 
\int \prod_{n,j}d{\bf r}_{jn}
\left( \prod_{n=1}^P \ {\rm det} (M^{(n)})\right)
\\ &\times& \nonumber
\exp\left(- \sum_{n,i<j} 
\frac{\tau\lambda}{|{\bf r}_{i,n}-{\bf r}_{j,n}|} \right), 
\end{eqnarray}
where the $N\times N$ matrix $M^{(n)}$ has the matrix elements
\[
M_{ij}^{(n)}=\langle{\bf r}_{i,n+1},\sigma_{i,n+1}|
e^{-\tau H_0}| {\bf r}_{j,n},\sigma_{j,n}\rangle.
\]
The last term in Eq.\ (\ref{partfun}) represents the Coulomb interaction 
between all $N$ electrons confined to the QD.
If there is no SO coupling, ${\rm det}M^{(n)}$ factorizes into 
a spin-up and a spin-down part, $S_z$ is a constant of motion, and 
the spin contamination problem arises. 
We mention in passing that the weight in the discretized path integral
(\ref{partfun}) is complex-valued, and one therefore may expect that
 observables  have an imaginary part. 
However, all statistical averages for physical observables
must have zero imaginary part, and this indeed we find within 
the standard stochastic error bars.
The discretized canonical many-particle partition function 
(\ref{partfun}) then allows us to access equilibrium
 observables of interest. 
For concreteness, we have chosen a rather low but finite 
temperature, $k_B T/\omega_0 = 0.1$.  Furthermore, unless
stated otherwise, simulations were carried out for $\alpha_D=0$ 
and interaction strength $\lambda=10$, which puts us into the
Wigner molecule regime. 
Note that by simply replacing $\alpha_R\to \alpha_D$,
results for $\alpha_R=0$ follow. 

\begin{figure}[t!]
\vspace{0.2cm}
\scalebox{0.3}{\includegraphics{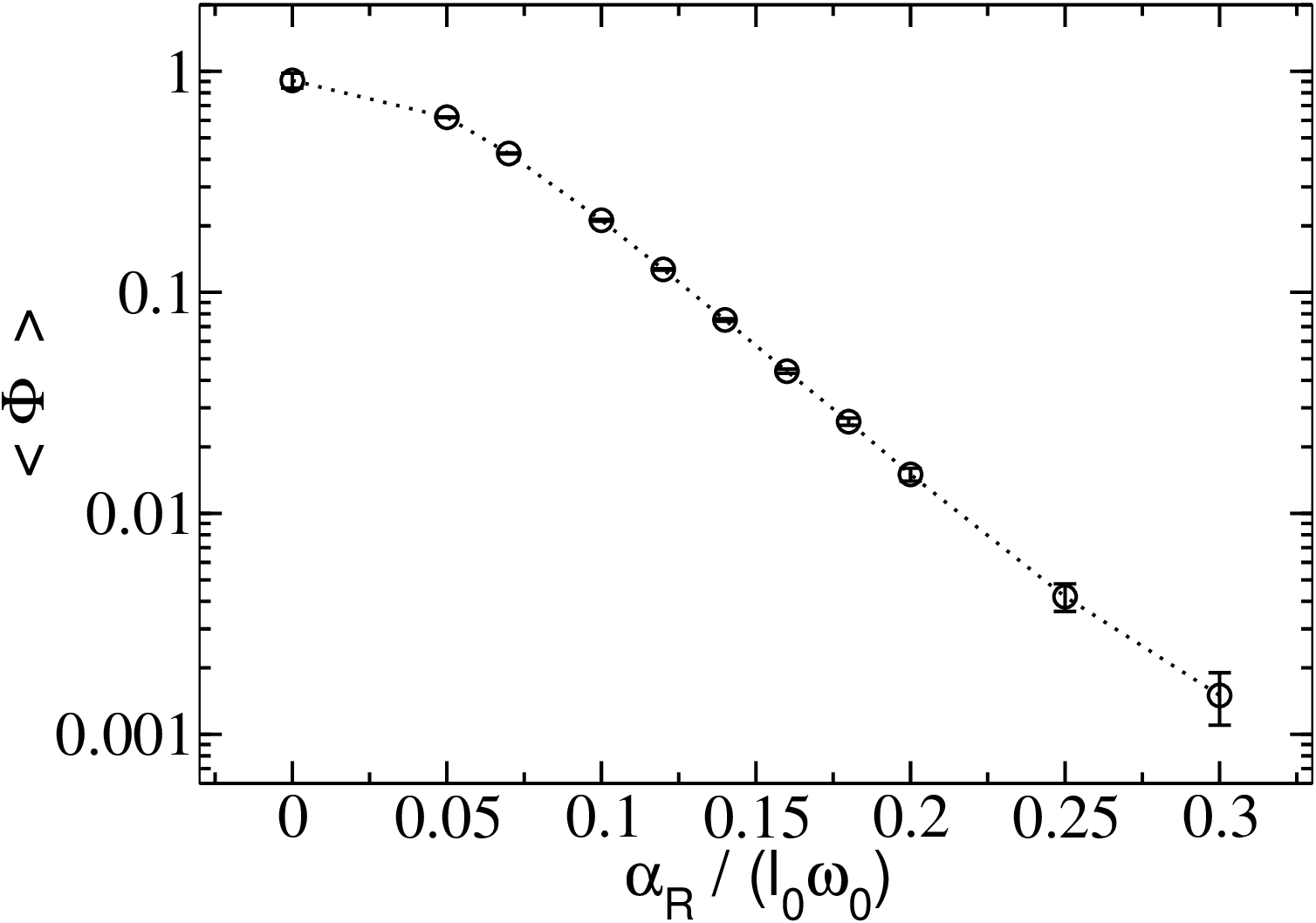}}
\caption{ \label{phivsa} 
Average sign $\langle\phi\rangle$ 
as a function of $\alpha_R$ (semi-logarithmic scale),
for $N=3, \alpha_D=0$ and $\tau\omega_0=0.25$.  The dotted curve is
a guide to the eye only.  Vertical bars denote standard
Monte Carlo error bars (one standard deviation).  }
\end{figure}

The main limitation for this type of PIMC simulation comes
from the fermionic 
sign problem. The sign problem generally arises when different paths that 
contribute to averages carry different signs, or even complex-valued phases,
as encountered in the case of non-zero SO couplings.
The resulting sign cancellation
 when sampling fermion paths then manifests itself as a very small 
signal-to-noise ratio.  
For instance, as a consequence of exchange,  these phases appear when forming
Slater determinants.  Here, in the presence of SO couplings, 
the sign problem occurs even for a single particle. 
Unfortunately, as a function of SO couplings, we find an 
exponential decay of the sign, see Fig.\ \ref{phivsa}.
This can be rationalized already on the single-particle level,
since the propagator acquires a complex phase factor in the presence
of SO couplings.  The propagator then resembles a real-time propagator
for a single particle, with time corresponding to the SO coupling.
For this problem, the exponential severity of the sign problem is well
established. In effect, the parameter 
regime where reliable simulations are possible  is limited to 
small-to-intermediate SO couplings. 
We note in passing that the Rashba SO coupling
$\alpha_R=(0.4-1.1)\times 10^{-11}eVm$
reported for the InGaAs dots of Ref.~\cite{Nitta1}
are about one order of magnitude larger than our largest value.
However, the SO coupling in InGaAs is also unusally strong, and our
values should apply more directly to GaAs dots.
 Simultaneously, the sign 
problem becomes more severe
when increasing $N$ or lowering temperature.
For the chosen parameters, the sign problem only allows to study QDs containing
$N<10$ electrons.  The average sign is $\langle \phi \rangle>0.001$ in
all cases reported below.  

We then compute several observables.  First, the energy of the 
$N$-electron dot can be obtained from
\begin{equation}\label{en}
E_N = -\frac{ \partial \ln Z(\beta)}{\partial \beta},
\end{equation}
where the derivative can be explicitly carried out
using Eq.~(\ref{partfun}).
Knowledge of the $E_N$ determines the addition energy
\begin{equation}
\Delta(N)=E_{N+1}-2E_N+E_{N-1}.
\end{equation}
A peak in the addition energy $\Delta(N)$ indicates 
 enhanced stability of the $N$-electron dot (``magic number'') \cite{reimann}.
Note that experimentally observed addition energies are determined by
{\sl free} energy differences, while we compute the energy.
However, for the low-temperature regime studied here, 
entropic contributions are smaller by about one order of magnitude,
and therefore our values for $\Delta(N)$ are of relevance to 
actual experiments.
Another quantity of interest is
 the total spin $S$, which we extract from the definition
\begin{equation}\label{s2}
\langle {\bf S}^2 \rangle = S(S+1), 
\end{equation}
where brackets denote the statistical average using Eq.~(\ref{partfun}),
and ${\bf S}$ is the total spin operator.
Finally, spatial ordering can be 
monitored via the electronic particle density, 
\begin{equation}
\label{chargeden}
n({\bf r})= \sum_{i=1}^N\langle\delta({\bf r}-{\bf r}_i) \rangle,
\end{equation}
which is normalized to $\int d{\bf r} n({\bf r})=N$. 
The related charge densities can also be accessed experimentally, e.g., via
capacitance spectroscopy or scanning tunneling microscopy 
techniques \cite{ashori}.   

As we work with finite discretization
$\tau$, Trotter approximation errors have to be taken 
into account \cite{deRaedt}. As shown in Ref.~\cite{Fye}, for small 
$\tau$, such errors vanish quadratically for all observables, allowing
for simple and efficient extrapolation schemes that completely
 eliminate this finite-$\tau$ error.  The $\tau^2$ scaling regime is 
reached for  $\tau \omega_0\leq 0.35$ when computing
the energy.  Furthermore, particle or spin densities  
are found to reach the $\tau^2$ scaling regime already at higher $\tau$. 
Results shown here have been carefully extrapolated down to $\tau=0$ 
with a linear regression fit, 
using results from several simulations obtained
 at $1/6 \leq \tau \omega_0 \leq 1/3$.
Hence no discretization errors are present, see also Ref.~\cite{Reusch},
and error bars denote just the standard stochastic Monte Carlo 
errors.

Spin flip moves are an essential
ingredient of the algorithm. Within the PIMC we allow for 
spin flips as well as for position moves. 
Single-particle moves were found sufficient to ensure ergodicity.
The average trial step size for position moves was 
adjusted to give acceptance rates of the order of $30$\%.
On the other hand, typical spin-flip acceptance rates were
much lower (several percent) and strongly dependent
on $\alpha_{R/D}$.  The possibility to change $S_z$
arises from the SO coupling and  can be used to circumvent
the spin contamination problem even for the case of no SO coupling,
namely by extrapolating finite-$\alpha_R$ results (where spins can
be flipped and no spin contamination problem is present) down to
$\alpha_R\to 0$.  Such a scheme yields 
the energy $E_N$ as well as the expectation value 
$\langle {\bf S}^2\rangle$ for the total spin of the many-body 
system at finite temperature. 
Typically, after $\approx 10^3$ equilibration passes, 
$1.5\times 10^7$  MC samples
were accumulated for each parameter set.
Our code runs at a speed of up to two weeks (for $N=9$, $P=60$) for a given
parameter set per $1.5\times 10^7$ samples on a standard 2 GHz Xeon processor.
We have checked our PIMC energies  for $N=2$ 
against finite-temperature exact diagonalization results, 
including both interactions and spin-orbit
couplings.  We found excellent agreement, validating
our approach.

\section{Results} \label{sec:Results}

\begin{figure}[t!]
\vspace{0.5cm}
  \scalebox{0.3}{\includegraphics{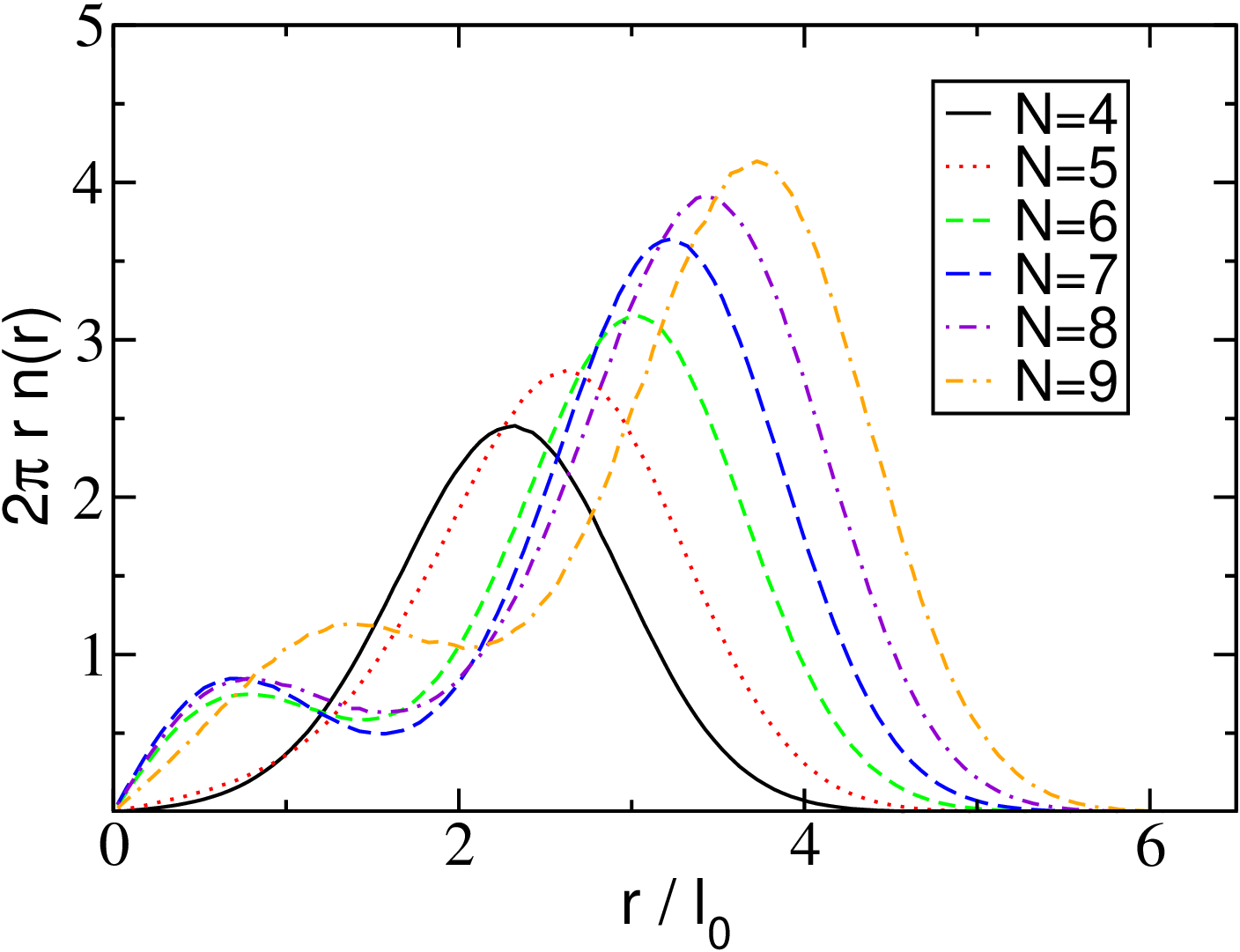} }
\caption{ \label{wigner} 
(Color online) Charge density for $N=4$ to $N=9$ electrons
at $\alpha_Rl_0=0.02$ and $\alpha_D=0$, $\lambda=10$ and $k_BT/\omega_0=0.1$.}
\end{figure}

To verify that for $\lambda=10$, we indeed have a Wigner molecule \cite{Egger},
let us start with the radially integrated particle density (\ref{chargeden})
shown in Fig.~\ref{wigner}. The cylindrical symmetry of the QD implies 
that $n({\bf r})$ only depends on the modulus  $r={\bf r}$. 
The plot indicates that the
sixth electron enters the center of the  
dot, whereas the remaining five electrons arrange on an outer ring in order
to minimize the Coulomb repulsion.  
More electrons are then added
to the outermost shell. Finally, for $N=9$, a second electron
enters the center.  This spatial shell filling sequence (as opposed
to orbital shell filling) is typical of the Wigner molecule, which forms
 the finite-size 
counterpart of a Wigner crystal.  In fact, precisely this spatial filling
sequence has been reported from a purely classical analysis (in 
particular, disregarding spin effects) \cite{lozovik,bedanov}.   
At higher temperatures, the Wigner molecule melts via thermal fluctuations,
while for lower $\lambda$, it is eventually destroyed by quantum 
fluctuations.    Generally, we find that particle densities are 
practically independent of the SO coupling strengths $\alpha_R$ 
or $\alpha_D$, at least for $\alpha_{R/D} l_0\leq 0.05$.

\begin{figure}[t!]
\vspace{0.2cm}
\scalebox{0.3}{\includegraphics{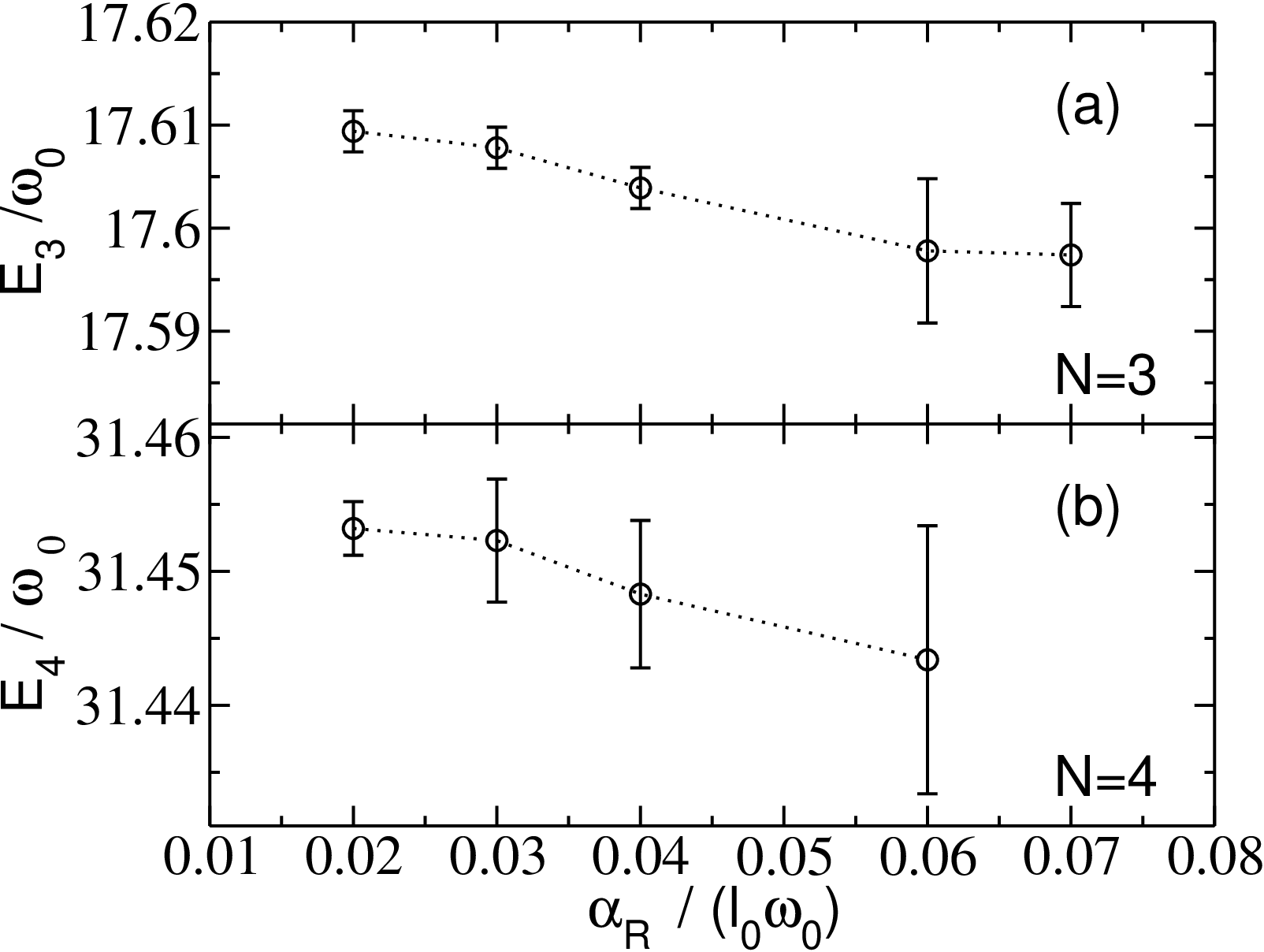} }
\caption{\label{ener34}
Energy $E_N$ in units of $\omega_0$
 for $(a)$ three and $(b)$ four electrons in the QD
as a function of $\alpha_R$ (here $\alpha_D=0$). 
 With increasing $\alpha_R$, the sign
problem becomes more severe, and thus the MC error 
bars tend to increase.}
\end{figure}

Although we consider rather small SO couplings, the many-body energy $E_N$
can be clearly seen to decrease 
as a function of $\alpha_R$ (here, $\alpha_D=0$). 
This trend has been observed in other studies as well \cite{destefani1}.
 In Fig.\ \ref{ener34}, we show this
effect for $N=3$  and $N=4$ electrons.

\begin{figure}[t!]
\vspace{0.2cm}
  \scalebox{0.3}{\includegraphics{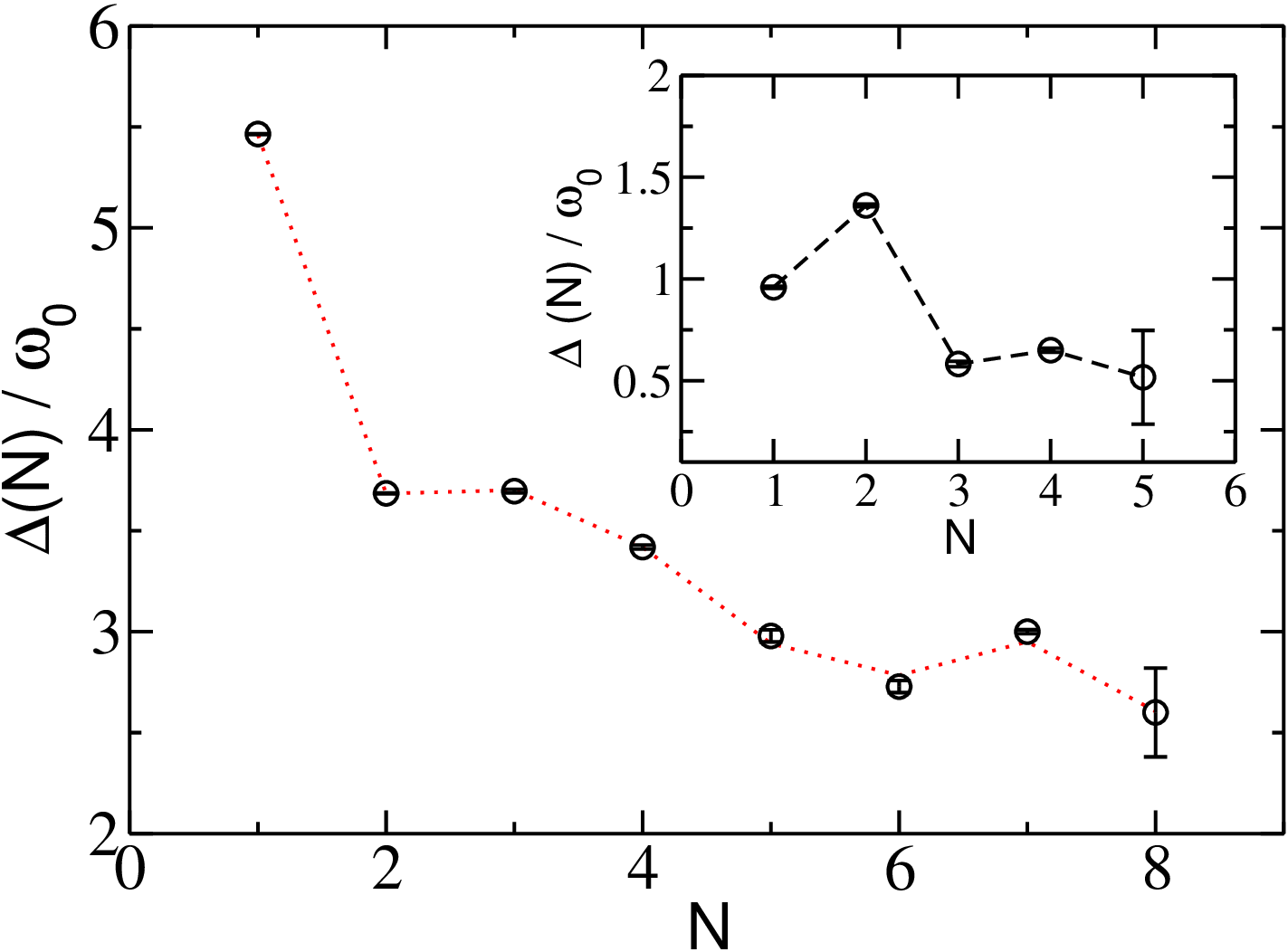} }
\caption{ \label{adden} (Color online)
Addition energy $\Delta(N)$ in units of $\omega_0$
for $\lambda=10$. Circles denote PIMC data extrapolated to $\alpha_R=0$. 
Note that $N=3$ represents a peak, since $\Delta(3)=3.696(9)$, 
whereas $\Delta(2)=3.685$ and $\Delta(4)=3.419(10)$. 
The dotted curve connects the corresponding PIMC results for 
$\alpha_R=0.04$. 
Inset:  Same for $\lambda=1$ and small $N$. The dashed curve is a guide to the
eye only.  }
\end{figure}

\begin{figure}[t!]
\vspace{0.2cm}
 \scalebox{0.3}{ \includegraphics{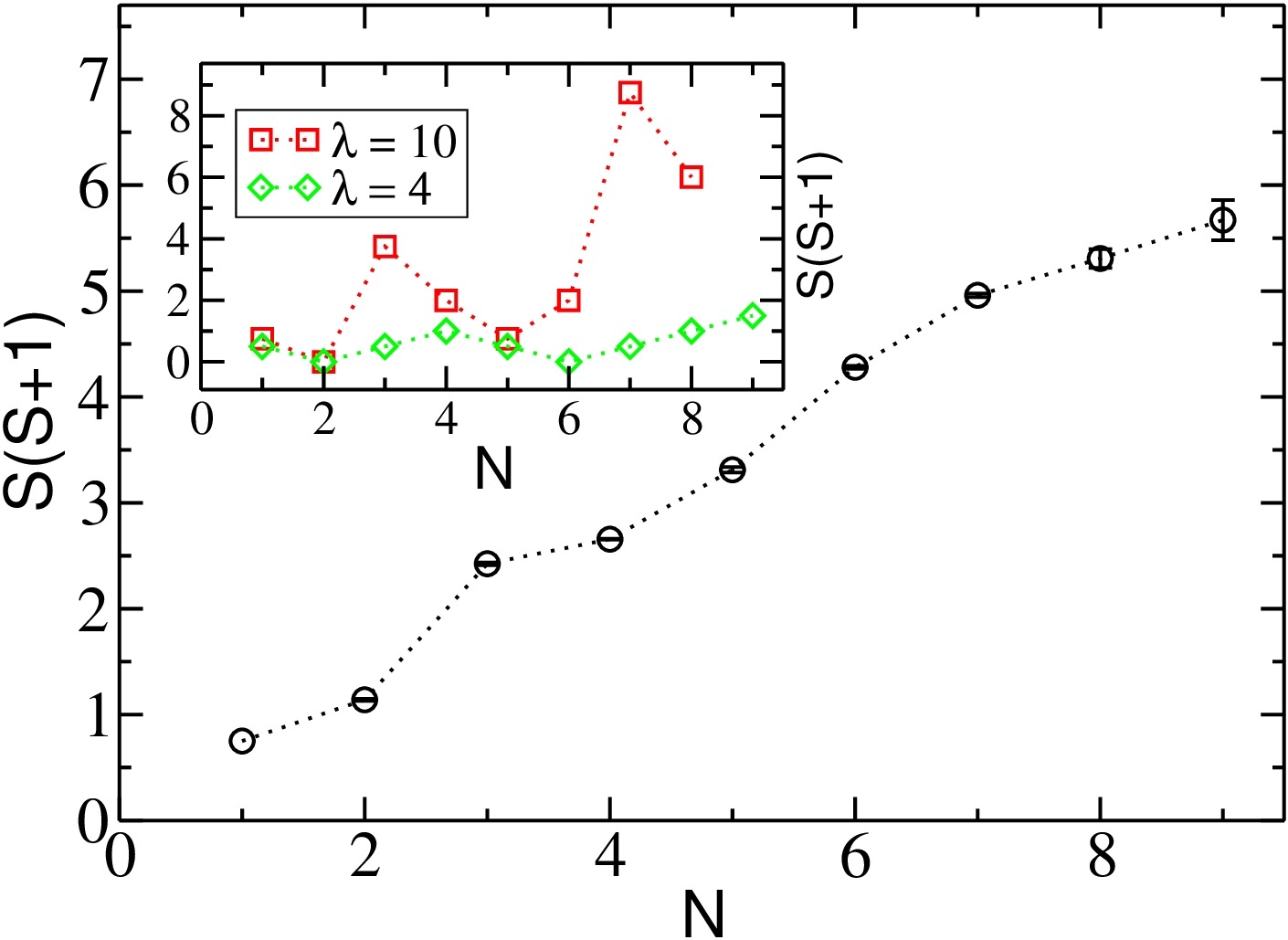}}
\caption{ \label{s2b} (Color online) $S(S+1)$ as defined in Eq.~(\ref{s2})
as a function of $N$ for $\lambda=10$ and
$k_B T /\omega_0=0.1$.  Dotted or dashed lines are guides to the eye only. 
Inset: Ground-state ($T=0$, $\alpha_{R/D}=0$) value for $S(S+1)$
from Ref.~\cite{Egger} for $\lambda=10$ (squares)
and for $\lambda=4$ (diamonds).
}
\end{figure}

For $\alpha_Rl_0\leq 0.07$, the SO coupling does not significantly 
influence addition energies for $N\leq 5$, see Fig.~\ref{adden}.
However, there is a slight increase in the addition energy $\Delta(6)$,
while the peak for $N=7$ is reduced for $\alpha_R=0.04$ (dotted curve).  
(In particular,
$\Delta(7)=3.00(1)$ for $\alpha_R\to0$, while $\Delta(7)=2.95(2)$ at 
$\alpha_R=0.04$.) 
Remarkably, the magic numbers $N=3$ and $N=7$ encountered in the Wigner 
molecule
regime are different from the ones for weak interactions (where a standard
Fermi liquid phase is present).  This is indicated in the inset 
of Fig.\ \ref{adden} for $\lambda=1$ and $\alpha_R=0.04$. 
The magic numbers for $\lambda=1$ can be rationalized in terms
of the subsequent filling of energy levels (orbitals) 
of a 2D harmonic oscillator. This predicts a peak for $N=2$, 
where the lowest level is filled, and another peak at $N=4$ reflecting
Hund's rule behavior, see also Ref.~\cite{governale}.
For strong Coulomb repulsion, 
the peak for $N=2$ is completely absent, 
while $N=3$  now corresponds to a magic number.
A very distinct peak in the addition energy is observed
for $N=7$.  To the best of our knowledge, this peak is not expected for weak 
interactions.
The picture of filling up spatial shells in the Wigner solid phase
(discussed above) may suggest that the 
filled {\it spatial} shell configurations 
$N=5$ and $N=8$ represent magic numbers. 
However, under a classical reasoning,
the addition energy should have no
pronounced peaks but exhibits a rather smooth and monotonic
 decay in $\Delta(N)$
\cite{lozovik,bedanov}.
Moreover, the classical prediction for $\Delta(N)$ \cite{bedanov}
yields values one order of magnitude smaller than the $\Delta(N)$ 
found here.  This indicates that for $\lambda=10$, despite
the clear onset of spatial ordering, 
quantum effects are still very important and cause the magic numbers 
$N=3$ and $N=7$. Notably, for these $N$, the  dot 
is seen to be partially spin-polarized.
The magic numbers for $N=3$ and $N=7$ can thus be rationalized
in terms of a Hund-rule  type behavior specific to
the incipient Wigner crystallized regime (see the discussion below and 
Fig.~\ref{s2b}). 
Our findings are therefore characteristic for the
quantum character of the Wigner molecule. 
A purely classical ``Wigner solid'' analysis \cite{lozovik,bedanov}
is expected to apply only for extremely large $\lambda$.
Our numerical PIMC results for $E_N$ (and the spin) as a function of $N$
and the Rashba coupling $\alpha_R$ 
 are summarized in Table \ref{valuetab}.
These data were all obtained for $\lambda=10$ and $k_B T/\omega_0=0.1$

\begin{figure}[t!]
\vspace{0.2cm}
  \scalebox{0.3}{\includegraphics{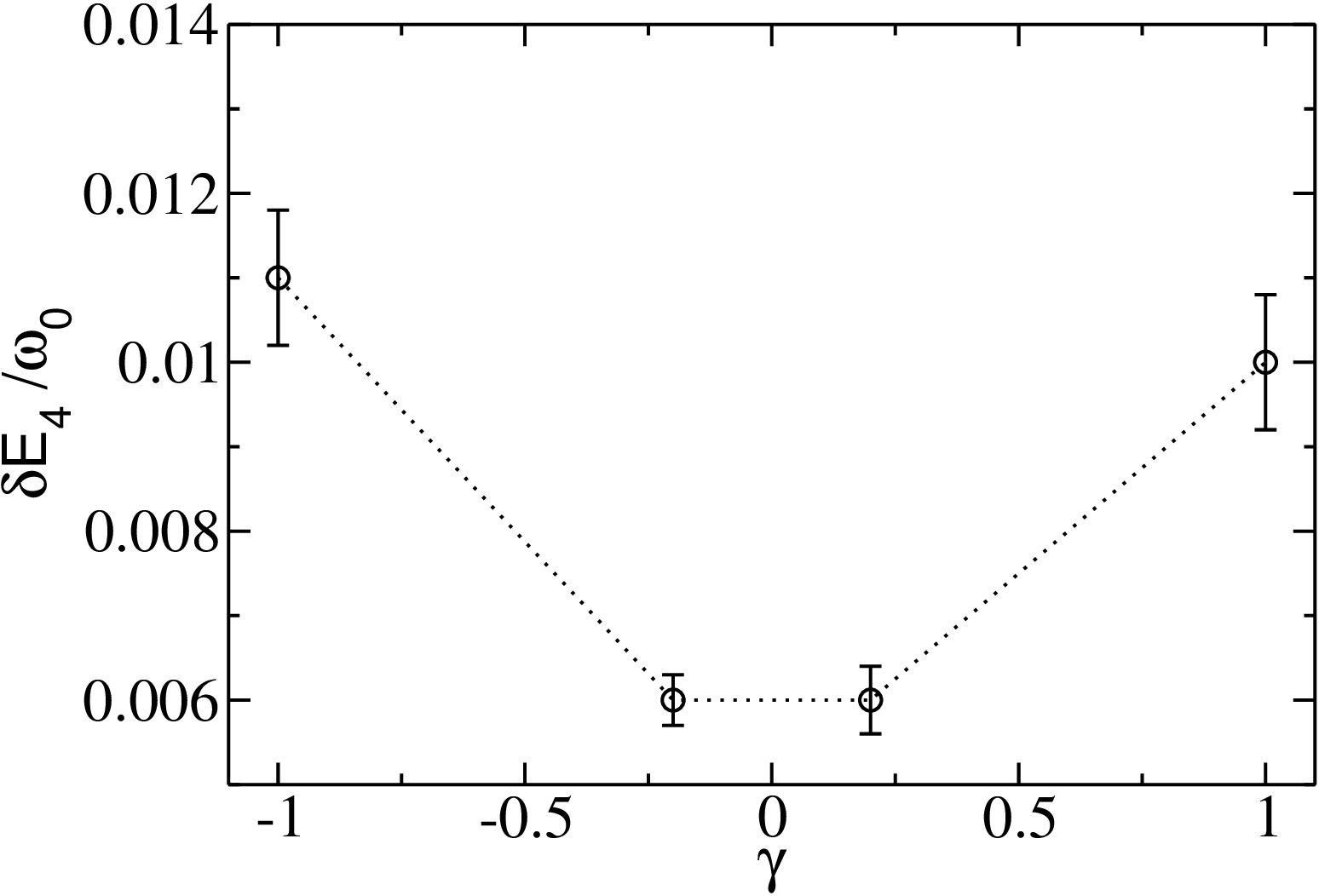} }
\caption{ \label{dEvsg} 
Energy difference
$\delta E_4=E_{(0,0)}-E_{(\alpha_R,\alpha_D)}$ for  $N=4$ and several
$\gamma$.  
The dotted line is a guide to the eye only.  }
\end{figure}

Next we consider the dependence of the energy on the two
types of spin couplings.  Let us take $N=4$ electrons,
again at $\lambda=10$, and fix $(\alpha_R+\alpha_D)l_0=0.05$.
We then study $E_4$ as a function of $\gamma=(\alpha_R-\alpha_D)/
(\alpha_R+\alpha_D)$, where $-1\leq \gamma \leq 1$ tunes
the relative strength of Rashba versus Dresselhaus coupling.
The result is depicted in Fig.\ \ref{dEvsg}, showing a symmetric curve 
$\delta E_4(\gamma)=E_{(0,0)}-E_{(\alpha_R,\alpha_D)}$. The symmetric shape 
can be explained by the unitary spin rotation which transforms the Rashba- 
into the Dresselhaus term and vice versa.  
With increasing $|\gamma|$, the relative weights of the Dresselhaus and
Rashba couplings increase, leading to increasing energy gains
$\delta E_4$  over the system without SO couplings.

Finally, we observe that spin plays a major role for $\lambda=10$.
Our  data for $S$ as defined in Eq.~(\ref{s2}) show a 
drastic influence of temperature. For such a strongly interacting system,
energies for different $S$ are typically close by, and a thermal average can 
then result in different $S$ as compared to ground-state results.
For instance, for $N=2$, we find a mixture of  
the singlet ground state $S=0$ and the excited triplet state $S=2$,
 as PIMC gives $S(S+1)=1.14(1)$ at $k_B T/\omega_0=0.1$.  
Due to the strong Coulomb interaction, 
there is a clear trend towards partial spin polarization
as a function of $\lambda$,  indicated in 
Fig.~\ref{s2b}, see also Table \ref{valuetab}.  
Moreover, the thermal average for $\langle {\bf S}^2 \rangle$
obtained from PIMC simulations at $\lambda=10$
increases monotonically with $N$, while the ground-state
spin \cite{Egger} has rather different
values and shows a nonmonotonic dependence on $N$. 
It is worth mentioning that the peaks in the addition energy for 
$N=3$ and $N=7$ correspond to enhanced values of $S$ as well.
Stability of the $N$ electron dot is thus connected to a tendency
towards spin polarization, reminiscent of Hund's rule behavior.

For $\lambda=4$, corresponding 
to weaker but still sizeable interactions, the dependence of
the ground-state spin \cite{Egger} is given in
the inset. The  shown nonmonotonic dependence on $N$ reflects the standard
Hund's rule physics. 
Comparing the $\lambda=4$ and $\lambda=10$ results, we observe
that interactions tend to further spin-polarize the dot.
The huge thermal effects observed
in the expectation value
$\langle {\bf S}^2\rangle$ also indicate that unless experiments
are carried out at extremely low temperatures, spin blockade phenomena
\cite{weinmann} relying on total spin selection rules will be thermally
washed out.   Finally, we note that for the SO couplings studied here,
spin expectation values were not significantly affected.

\section{Conclusion}
\label{conc}

We have investigated the behavior of up to nine electrons in a 
quantum dot. We took into account strong Coulomb correlations between 
the electrons, and also incorporated spin-orbit couplings.  Our results
were obtained from 
path-integral quantum Monte Carlo simulations.
 An exponential decrease of the fermionic sign is found with increasing
SO couplings. This sign problem appears even for a 
single electron. 
Nevertheless, simulations are possible for weak SO couplings,
where their inclusion can also be used to eliminate the 
spin contamination problem.

We observe peaks in the addition energy 
spectrum for $N=3$ and $N=7$, which are likely to 
correspond to the stability of partially spin-polarized configurations
induced by Coulomb interactions.  These peaks are neither expected
in the weak-interaction regime nor in the classical (deep) Wigner 
solid, where spin effects are negligible.  We hope that this
prediction can soon be tested experimentally. 
Our results were obtained in the regime of weak  spin-orbit couplings,
since otherwise numerical instabilities associated with the sign problem
occur.  Given this restriction however, PIMC offers a powerful tool
to analyze the effects of spin-orbit couplings in strongly 
interacting quantum dots.  We find no dramatic effects, but observable
downward shifts in the many-body energy that scale quadratically in the
spin-orbit couplings.  Spin-orbit couplings also affect addition energies.

\acknowledgments
This work was supported by the DFG under the Gerhard-Hess program. We thank 
M. Thorwart for a careful reading of the manuscript.

\newpage
\eject

\hspace{1cm}
\begin{table}
\begin{tabular}{|ll|l|l|}\hline
$N$&$\alpha_Rl_0 $&$E/\omega_0$&$S(S+1)$\\\hline\hline
$1$&$0$         &     $0.9988(1)$  & $0.75(0)$\\
$1$&$0.04$        &     $0.9986(6)$  & $0.75(0)$\\\hline
$2$&$0$         &     $7.464(2)$   & $1.14(1)$\\
$2$&$0.04$        &     $7.459(3)$   & $1.11(2)$\\\hline
$3$&$0$         &     $17.610(1)$  & $2.424(13)$\\
$3$&$0.04$        &     $17.603(2)$  & $2.426(16)$\\\hline
$4$&$0$         &     $31.454(1)$  & $2.657(4)$  \\
$4$&$0.04$        &     $31.448(6)$  & $2.654(31)$ \\\hline
$5$&$0$         &     $48.717(1)$  & $3.312(26)$ \\
$5$&$0.04$        &     $48.712(14)$ & $3.339(71)$\\\hline
$6$&$0$         &     $68.959(1)$  & $4.280(13)$  \\
$6$&$0.04$        &     $68.917(30)$ & $4.27(12)$ \\\hline
$7$&$0$         &     $91.929$     & $4.96(18)$ \\
$7$&$0.04$        &     $91.906(30)$ & $4.89(11)$\\\hline
$8$&$0$         &     $117.889$    & $5.307(89)$  \\
$8$&$0.04$        &     $117.83(6)$  & $5.37(35)$  \\\hline
$9$&$0$         &     $146.501(1)$ & $5.67(19)$\\
$9$&$0.04$        &     $146.36(22)$ & $5.95(68)$\\\hline
\end{tabular}
\caption{PIMC data for $E_N$ and $S(S+1)=\langle {\bf S}^2\rangle$
as a function of $\alpha_R$ for $\alpha_D=0, \lambda=10,$ and 
$k_B T/\omega_0=0.1$. The $\alpha_R=0$ data are taken from 
extrapolations. Bracketed numbers denote error estimates.}
\label{valuetab}
\end{table}

\newpage

\end{document}